\documentclass[%
reprint,
nofootinbib,
 amsmath,amssymb,
 aps,
]{revtex4-1}
\usepackage{epsfig,epsf,rotating,wrapfig,tabularx}
\usepackage{dcolumn}
\usepackage{bm}
\usepackage{graphicx}

\begin{document}

\preprint{APS/123-QED}

\title{An analysis of charged particles spectra in events with different charged multiplicity}

\author{\firstname{A.~A.}~\surname{Bylinkin}}
 \email{alexander.bylinkin@desy.de}
\affiliation{%
 Institute for Theoretical and Experimental
Physics, ITEP, Moscow, Russia
}%
\author{\firstname{A.~A.}~\surname{Rostovtsev}}
 \email{rostov@itep.ru}
\affiliation{%
 Institute for Theoretical and Experimental
Physics, ITEP, Moscow, Russia
}%


\begin{abstract}
The shapes of invariant differential cross section for charged particle production as function of transverse momentum measured in $pp$ collisions by the STAR detector are analyzed. The spectra shape varies with the event charged multiplicity changing. To describe this and several other recently observed effects a simple qualitative model for hadroproduction mechanism was proposed.
\end{abstract}

\pacs{Valid PACS appear here}
\maketitle


Inclusive charged particle distributions have been studied for a long time to derive the general properties of hadronic interactions at high energies. A large body of the experimental data on charge particle production spectra in baryon-baryon, gamma-baryon and gamma-gamma collisions has been accumulated during last forty years. However, the underlying dynamics of hadron production in high energy particle interactions is still not fully understood. A comparative detailed analysis of the measured spectra of charged particles produced in different type of collisions could shed light on the hadroproduction mechanisms.   

Recently, a new unified approach to describe the particle production spectra shape was proposed~\cite{OUR1}. It was suggested to approximate the charged particle spectra as function of the particle’s transverse momentum~($P_T$) by a sum of an exponential (Boltzmann-like) and a power law statistical distributions. 
\begin{equation}
\label{eq:exppl}
\frac{d\sigma}{P_T d P_T} = A_e\exp {(-E_{Tkin}/T_e)} +
\frac{A}{(1+\frac{P_T^2}{T^{2}\cdot n})^n},
\end{equation}
where  $E_{Tkin} = \sqrt{P_T^2 + M^2} - M$
with M equal to the produced hadron mass. $A_e, A, T_e, T, n$ are the free parameters to be determined by fit to the data.  The detailed arguments for this particular choice are given in~\cite{OUR1}.  

\begin{figure}[h]
\includegraphics[width =8cm]{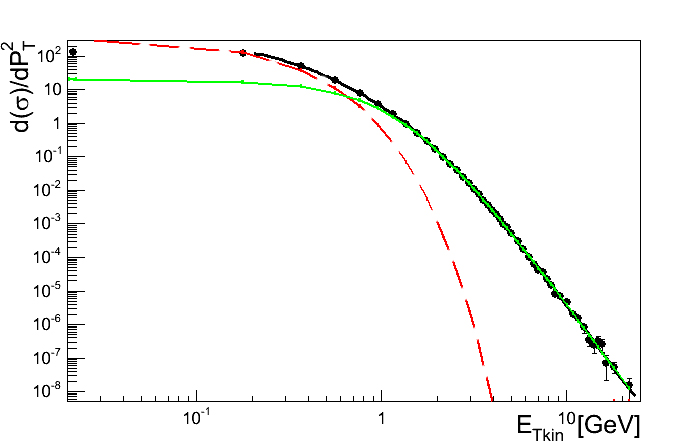}
\caption{\label{fig:0} Charge particle differential cross section~\cite{UA1} fitted to the function~(\ref{eq:exppl}): the red (dashed) line shows the exponential term and the green (solid) one - the power law.}
\end{figure}

\begin{figure*}[!]
\includegraphics[width =18cm]{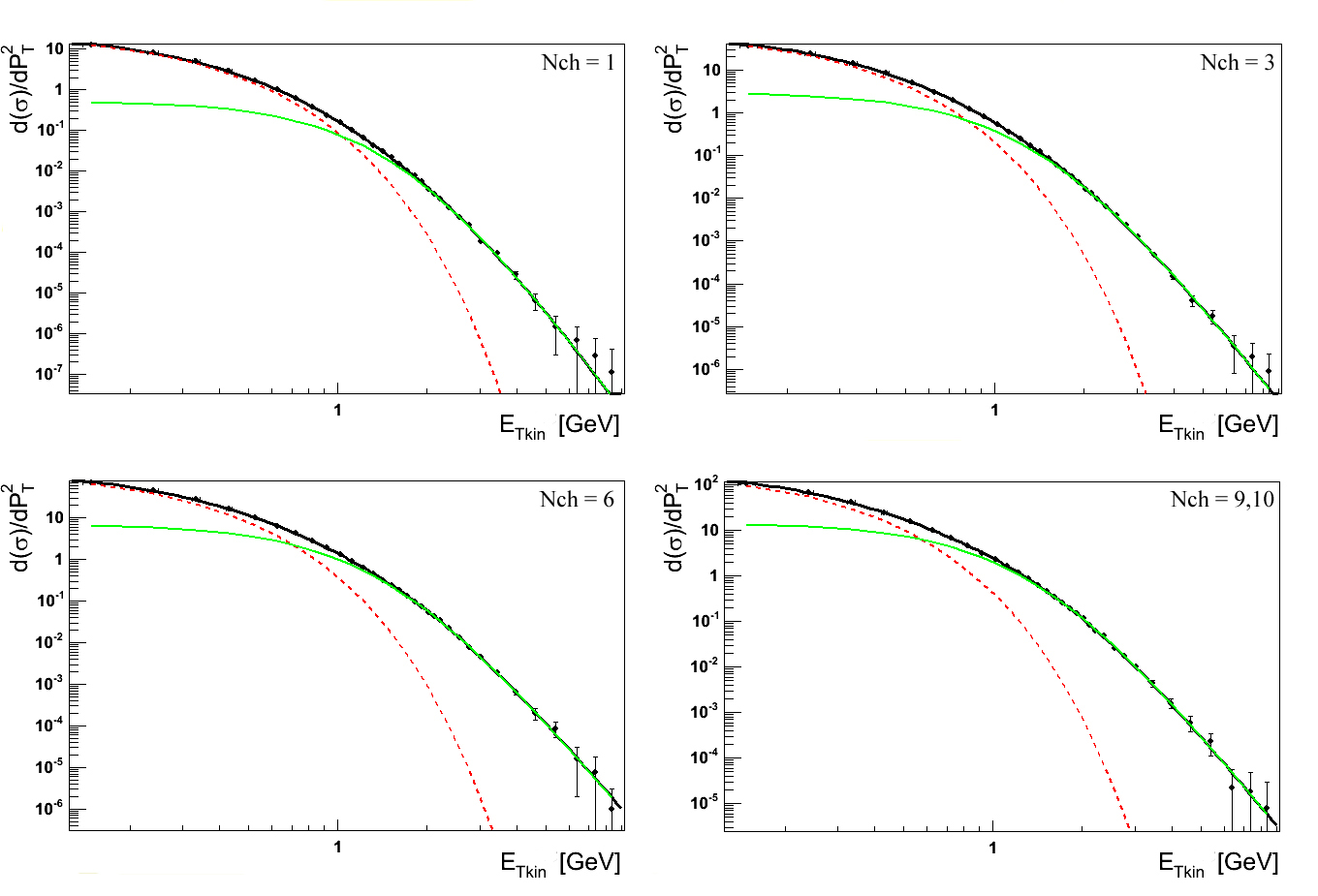}
\caption{\label{fig:01} Charged particle spectrum~\cite{STAR} for different values of charged multiplicity fitted to the function~(\ref{eq:exppl}): the red (dashed) curve shows the exponential term and the green (solid) one stands for the power law term.}
\end{figure*}

The proposed new parameterization is represented by a sum of an exponential and a power law functional terms. The variations of the parameters of this approximation were studied as function of energy and type of colliding particles, as well as of other experimental conditions.
A typical charged particle spectrum as function of transverse energy, fitted with this function~(\ref{eq:exppl}) is shown in Fig~\ref{fig:0}. 

The contributions of the exponential and power law
terms
of the parameterization~(\ref{eq:exppl}) to
the typical spectrum of charged particles produced in $pp-$collisions are also shown separately in Figure~\ref{fig:0}.
The relative contribution of these terms is characterized by ratio $R$ of the power law term alone to the parameterization function integrated over $P_T^{2}$:

\begin{equation}
R = \frac{AnT}{AnT + A_e(2MT_e + 2T_e^2)(n-1)}
\end{equation}

It was found that the exponential contribution dominates the charged particle spectra in baryon-baryon collisions while it is completely missing in gamma-gamma interactions~\cite{OUR1}. Moreover, the exponential contribution is characteristic for charged pion production and is much less pronounced in kaon or proton (antiproton) production spectra~\cite{OUR2}. There is also no room for the exponential contribution in a heavy quarkonium production in pp collisions~\cite{CDF}. 

In order to explain this observed phenomenon it was suggested to consider two distinct sources of the hadrons produced in particle-particle collisions. One is a radiation of hadrons by the preexisted valence quarks. This source of hadrons is characteristic for colliding baryons and completely missing for colliding gamma quanta.  Another source of hadrons is related to QCD-vacuum fluctuations, described by the Pomeron exchange. This source of particles exists in any hadronic interaction, including $\gamma\gamma$ collisions. If so, one could relate the exponential (Boltzmann-like) part of the whole particle spectrum to a quasithermalized radiation from the valence quarks\footnote{In the framework of the non-relativistic quark model the $P_t$ distribution of valence quarks have usually the Gaussian form}, while Pomeron interactions give rise to the hadrons distributed according a power law statistics\footnote{Perturbative QCD Pomeron produces the (mini)jets with a power-like $k_t$ distribution.}. 

Although this picture is a qualitative one, it allows making several predictions:

1.	In $pp$-collisions the QCD-fluctuations are more flavor democratic with respect to the valence quark related radiation. This results in much smaller exponential contribution to the charged kaon spectra produced in baryon-baryon interactions then that to the charged pion spectra. 

2.	The AGK cutting rules~\cite{AGK} state that charge multiplicity in hadronic interactions is proportional  to the number of Pomerons involved in this interaction. Therefore, the relative contribution of the exponential part of the approximation (\ref{eq:exppl}) will decrease with the increase of charged multiplicity in $pp$ interactions.

3.	In high energy photon-proton collision the exponential contribution to the charged particle spectra will strongly depend on rapidity of produced charged particles. The hadrons produced in the proton direction in rapidity space will show a sizable exponential contribution to their spectra, while the distribution of hadrons produced on the photon side of the event will be described by the power law function only. Therefore, one expects to observe a change between these two regimes for particles produced around zero rapidities in the photon-proton center of mass system. 

4.	Another prediction could be made about an absolute dominance of the exponential contribution to the spectra of particles produced in the high rapidity proton fragmentation region where the role of valence quark is more important.

\begin{figure}[h]
\includegraphics[width =8cm]{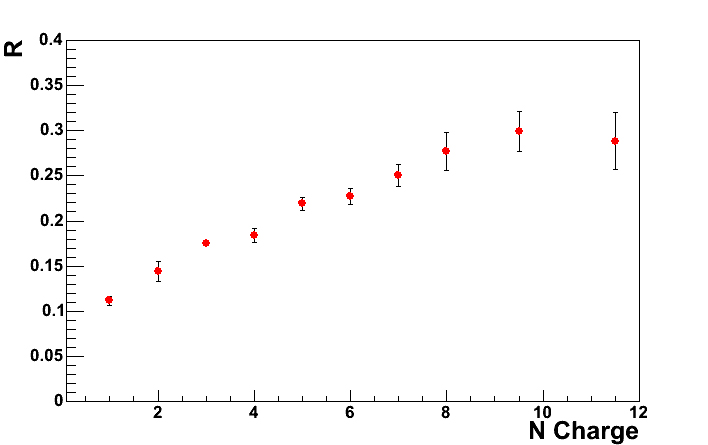}
\caption{\label{fig:02} The relative value of the power law term contribution into the approximation (\ref{eq:exppl}) as function of the visible charge event multiplicity.}
\end{figure}

The prediction (1) about the flavor democracy in the Pomeron interactions was recently proved to be true by the comparative analysis of the data provided by the PHENIX experiment for $pp$ collisions at RHIC~\cite{Adare:2011vy}.  Moreover, it turned out that the difference in the relative contribution of the exponential part in the approximation (\ref{eq:exppl}) for pions and kaons defines the peculiar shape of the kaon to pion yield ratio as function of the hadron transverse momentum.   
The prediction (3) could be checked with a detailed measurement of the photon-proton interaction at HERA experiments. For that the double differential distribution of produced charged particles in transverse momentum and rapidity space has to be measured in the photon-proton center of mass system. In order to check the prediction (4) one needs to run a dedicated experiment able to measure charged particle spectra at very high rapidities. 
To check the prediction (2) it is possible to use already available data published by the STAR experiment~\cite{STAR}. These data are represented by charged particle spectra for the $pp$-collision events with different visible charged multiplicity. In the STAR experiment the visible charged multiplicity and charged particle spectra were measured in the limited central rapidity interval ($-0.5<y<0.5$). Four examples of these spectra together with the fit to the function (\ref{eq:exppl}) are shown in Figure \ref{fig:01}. It can be seen from this figure that the relative contribution of the exponential part is decreasing with the visible charged event multiplicity rising as it is expected from the point of view of the simple model given above. The dependence of the exponential contribution in the approximation (\ref{eq:exppl}) needed to describe the spectrum shape of the produced charged particles as function of the visible charged event multiplicity is shown in Figure \ref{fig:02}. The functional trend demonstrated in Figure \ref{fig:02} is in accord of the qualitative predictions discussed above. However, this trend is largely smeared by the multiplicity fluctuations due to very limited rapidity window available for the measurement in the STAR experiment. In order to obtain a stronger non-smeared dependence on charged multiplicity one has to make similar measurements over whole available event rapidity space.

In conclusion, a simple naive model for hadroproduction mechanism was proposed to explain a number of recently observed phenomena. Within the framework of this model there are two distinct sources of hadrons produced in particle-particle collisions. One is a radiation of hadrons by the preexisted valence quarks. This source of hadrons is characteristic for colliding massive baryons and is completely missing for colliding gamma quanta.  Another source of hadrons is related to QCD-vacuum fluctuations, described by the Pomeron exchange. The Pomeron interactions give rise to the hadrons distributed according the QCD-like power-law statistical distribution, while the valence quark radiation results in a formation of the exponential Boltzmann-like spectrum of produced particles. The predicted within this model decrease of the relative exponential contribution to the approximation (\ref{eq:exppl}) describing the particle spectra in $pp$-collisions with rising event charged multiplicity was proved on the data previously published by the STAR Collaboration.

The authors thank Professor M.G.Ryskin for fruitful discussion and his help provided during the preparation of this short note.


\end{document}